\documentstyle[tighten,preprint,pra,aps]{revtex}
\begin{document}
\draft
\title
      {Generalized Relativistic Effective Core Potential Method:\\
       Theory and calculations}

\author{A.\ V.\ Titov\cite{Email} and N.\ S.\ Mosyagin}
\address{St.-Petersburg Nuclear Physics Institute,\\
         Gatchina, St.-Petersburg district 188350, RUSSIA}

\date{\today}
\maketitle

\begin{abstract}
 In calculations of heavy-atom molecules with the shape-consistent
 Relativistic Effective Core Potential (RECP), only valence and some outer-core
 shells are treated explicitly, the shapes of spinors are smoothed in the
 atomic core regions and the small components of four-component spinors are
 excluded from calculations. Therefore, the computational efforts can be
 dramatically reduced. 

 However, in the framework of the standard
 nodeless
 radially local RECP versions, any
 attempt to extend the space of explicitly treated electrons more than some
 limit does not improve the accuracy of the calculations. The errors caused by
 these (nodeless) RECPs can range up to 2000 $cm^{-1}$ and more for the
 dissociation and transition energies even for lowest-lying excitations that
 can be unsatisfactory for many applications. Moreover, the direct calculation
 of such properties as electronic densities near heavy nuclei, hyperfine
 structure, and matrix elements of other operators singular on heavy nuclei is
 impossible as a result of the smoothing of the orbitals in the core regions.

 In the present paper, ways to overcome these disadvantages of the RECP method
 are discussed. The developments of the RECP method suggested by the authors
 are studied in many precise calculations of atoms and of the TlH, HgH
 molecules. The technique of nonvariational restoration of electronic structure
 in cores of heavy atoms in molecules is applied to calculation of the
 $P,T$-odd spin-rotational Hamiltonian parameters including the weak
 interaction terms which break the symmetry over the space inversion ($P$) and
 time-reversal invariance ($T$) in the PbF, HgF, BaF, and YbF molecules.

\vspace{1cm}
{\bf SHORT NAME:} GRECP method: Theory and calculations.

{\bf KEYWORDS FOR INDEXING:} Relativistic Effective Core Potential
    (Pseudopotential), Ab initio relativistic method,
    Electronic structure calculation, Molecules with heavy atoms.

\end{abstract} 

\section{Introduction}
 \label{Intro}

 The Dirac-Coulomb (DC) Hamiltonian which is used for calculation of heavy
 atoms and heavy-atom molecules has the form ($e = m_e = \hbar = 1$):

\begin{equation}
     {\bf H}^{DC}\ =\ \sum_{i} {\bf h}^D(i)
                    + \sum_{i>j} \frac{1}{r_{ij}} \ ,
 \label{DC}
\end{equation}
 where the one-electron Dirac operator ${\bf h}^{D}$ is
\[
   {\bf h}^D\ =\ -\imath c ({\vec \alpha} \cdot {\vec \nabla})
                 + c^2 (\beta - 1) + V_{nuc} \ ,
\]
 ${\vec \alpha}, \beta$ are the $4\times 4$ Dirac matrices, $c$ is the speed
 of light and $V_{nuc}$ is the nuclear potential which can take account of
 effect of finite nuclear size etc.

 Higher approximation levels involve modifying the instantaneous Coulomb
 interaction between electrons. These modifications are derived from the
 Quantum ElectroDynamic (QED) theory. The simplest approximation is obtained by
 including the one-photon exchange. In the Coulomb gauge it leads to so-called
 Dirac-Coulomb-Breit (DCB) Hamiltonian (see, e.g., \cite{Lindgren})

\begin{equation}
    {\bf H}^{DCB}\ =\ {\bf H}^{DC} + B_{ij}\ ,
 \label{DCB}
\end{equation}
 where

\[
   B_{ij}(\omega_{ij})\ =\
          \sum_{i>j} [-({\vec \alpha}_i \cdot {\vec \alpha}_j)
          \frac{\cos (\omega_{ij} r_{ij})}{r_{ij}}
        + ({\vec \alpha}_i \cdot {\vec \nabla}_i)
          ({\vec \alpha}_j  \cdot {\vec \nabla}_j)
        \frac{\cos (\omega_{ij} r_{ij}) - 1}{\omega_{ij}^2 r_{ij}}]\ ,
\]
 and $\omega_{ij}$ denotes the transition frequency between the electrons
 $i$ and $j$.

 A low-frequency expansion of the cosines yields the incomplete Breit
 interaction $B_{ij}(0)$:

\[
   B_{ij}(0)
             = -\frac{1}{2}[{\vec \alpha}_i \cdot {\vec \alpha}_j
              + ({\vec \alpha}_i \cdot {\vec r}_{ij})
                ({\vec \alpha}_j \cdot {\vec r}_{ij})/r_{ij}^2]/r_{ij} \ .
\]
 These terms describe the instantaneous magnetostatic interaction and
 retardation in the electric interaction between electrons.

 In calculations on heavy-atom molecules, the DC and DCB Hamiltonians are
 usually replaced by an effective Hamiltonian

\begin{equation}
  {\bf H}^{Ef}\ =\ \sum_{i_w} [{\bf h}^{Schr}(i_w) + {\bf U}^{Ef}(i_w)]
              + \sum_{i_w > j_w} \frac{1}{r_{i_w j_w}} \ ,
 \label{Hef}
\end{equation}
 written only for valence and some outer core electrons denoted by indices
 $i_w$ and $j_w$; ${\bf U}^{Ef}$ is an Relativistic Effective Core Potential
 (RECP) operator simulating interactions of the explicitly treated
 (``pseudo-valence'') electrons with those which are excluded from the RECP
 calculation. In Eq.~(\ref{Hef}), ${\bf h}^{Schr}$ is the nonrelativistic
 one-electron Schr\"odinger operator

\[
     {\bf h}^{Schr}\ = - \frac{1}{2} {\vec \nabla}^2 + V_{nuc} \ ,
\]
 Contrary to the four-component (relativistic) wave function used in DC(B)
 calculations, the pseudo-wave function in the RECP case can be both two- and
 one-component.

\section{Features of shape-consistent RECPs}
 \label{ShC-RECPs}

 The RECP approximations with the radially local operator for
 ``shape-consistent'' (or ``norm\--con\-ser\-ving'') pseudoorbitals
 (pseudospinors)\ \cite{Durand,Christ,Lee} are the most widely used in modern
 calculations of molecules with heavy elements. Below we discuss mainly these
 RECP versions and their developments because they provide the largest
 computational savings for the same level of accuracy as compared with other
 RECP versions.

\subsection{Advantages}
 \label{Advantages}

\begin{itemize}
\item Chemically inactive electrons are excluded from RECP calculations.

\item The valence orbitals (spinors) are smoothed in the core regions
      of heavy atoms to generate nodeless pseudoorbitals (pseudospinors).
 Therefore, the number of the one-electron basis functions may be minimized,
 thus reducing dramatically both the number of two-electron integrals and the
 computational time.

\item The small components of four-component spinors are eliminated and the
      nonrelativistic kinetic energy operator is used.
 The RECP method allows one to use a well-developed nonrelativistic technique
 of calculation and relativistic effects are taken into account with the help
 of spin-dependent semilocal potentials. As a result, the most part of
 difficulties inherent for the DC(B) molecular calculations can be avoided.

\item In principle, correlations of the explicitly treated electrons with
      those which are excluded from the RECP calculation can be considered
      within ``correlated'' RECP versions.
 Reducing the number of explicitly correlated electrons with the help of
 the correlated RECPs is a very promising way to minimize efforts when
 performing high-precision molecular calculations.

\end{itemize}

\subsection{Disadvantages}
 \label{Disadvantages}

\begin{itemize}
\item By now, different schemes of the RECP generation are suggested which
      use the radially local operator for the effective potential. However, as
      it was demonstrated in many test calculations, they provide the comparable
      level of accuracy for the same number of the explicitly treated
      electrons.

      It is clear that the explicit inclusion of the outer core electrons
      into the RECP calculation is the way to increase the computational
      accuracy. However, the extension of the space of these electrons more
      than some limit does not improve the accuracy
      as is obtained in all our atomic calculation with RECPs and PPs.
      Although the
      errors caused by the nodeless RECP approximations can be small enough in
      competent treatments, they still range up to 2000 $cm^{-1}$ and more
      for the dissociation and transition energies even for lowest-lying
      excitations that can be unsatisfactory in many applications.

      One can conclude that the scheme of smoothing and the number of the
      explicitly treated electrons are not only responsible for refining the
      accuracy of the RECP calculations.

\item The reliability of the radially local RECP versions is not high for
      transitions with excitations from the outer core shells, for the
      chemistry of transition, rare-earth elements, etc.


\item Moreover, the direct calculation of such properties as electronic
      densities near heavy nuclei, hyperfine structure, and matrix elements
      of other operators singular on heavy nuclei is impossible as a result
      of the smoothing of the orbitals in the core regions of heavy elements.

\end{itemize}

 To overcome these disadvantages of the shape-consistent RECP method, a few
 developments were done by the authors concerning the RECP operator first of
 all~\cite{Titov1,Tupitsyn1,SfC-RECP,Comment1,GRThGr}. A detailed
 theoretical analysis of the RECP method is presented in~\cite{GRThGr}.

\subsection{Historical background of RECP method}
 \label{Background}

 Some of the most important achievements in the development of the effective
 core potential (or pseudopotential) theory are listed below:

\begin{itemize}
 \item The first ideas and papers devoted to the PseudoPotential (PP) theory of
 Hellmann (1935)~\cite{Hellmann} and Gombas (1935)~\cite{Gombas} were published more than 60 years ago.

\item The procedure of treatment of the valence electrons in the frozen sea
 of core electrons was suggested by Fock {\it et al.}\ \cite{Fock} in 1940.

\item The smoothing of orbitals in atomic cores was suggested and
 the orbital angular momentum ($l$) dependence of the PP operator for the
 case of pseudoorbitals smoothed in the cores was first emphasized in the
 pioneering paper of Phillips \& Kleinman (1959)~\cite{Phillips}.

\item The PP operator with the angular projectors was used first by Heine and
 Abarenkov (1964, 1965)\ \cite{Abarenkov} to generate one-electron model
 potentials mainly for the solid-state calculations.

 \item Improvement of the PP accuracy when treating several valence electrons
 explicitly was suggested by Weeks \& Rice \cite{Weeks} in 1968.

\item Goddard~III suggested a scheme (1968)\ \cite{Goddard} for the generation
 of the radially local effective potentials on the base of inverting the atomic
 Hartree-Fock (HF) equations.

 \item The ab initio radially local effective potentials were efficiently
 applied by Kahn {\it et al.} (1976)~\cite{Kahn} to molecular calculations.

\item The nonrelativistic shape-consis\-tent ECP was suggested by Durand \&
 Barthelat (1975)\ \cite{Durand}. Christiansen {\it et al.} (1977)\
 \cite{Christ} have modified their smoothing scheme.

\item The Relativistic ECP method for two-component atomic pseudospinors was
 proposed by Lee {\it et al.} (1977)\ \cite{Lee}.

\item Hafner \& Schwarz (1979)\ \cite{Hafner} have split the RECP operator on
 the spin-independent and spin-orbit parts that allowed one to take into
 account the spin-orbit interaction only at the final stage of a calculation
 thus reducing computational efforts.

\end{itemize}

\subsection{Notations}
 \label{Notation}

 Below we shall use the following notations for the principal quantum numbers
 $n$ of orbitals: $n_f$ for the inner core (IC) orbitals, $n_c$ for the
 outer core (OC) orbitals, $n_v$ for the valence (V) orbitals, and $n_a$ for
 the virtual (A) orbitals. The quantum numbers of the orbital momentum, total
 momentum, and its projection will be designated by the $l$, $j$, and $m$
 indices, respectively.  We shall write one-electron states as

\[
  \widetilde{\psi}_{nljm}(\tau) =
   \widetilde{\varphi}_{nlj}(r) \chi_{ljm}(\Omega,\sigma)
\]
 for the pseudospinors and as

\[
\psi_{nljm}(\tau) =
   \left(
      \begin{array}{ll}
       P_{nlj}(r) & \chi_{ljm}(\Omega,\sigma) \\
       Q_{nlj}(r) & i\chi_{l'jm}(\Omega,\sigma)
      \end{array}
   \right)
\]
 for the four-component spinors where

\[
  l'=2j-l, \;\;\; \tau \equiv (\vec{r}, \sigma) \equiv (r,\Omega,\sigma),
\]
 $\widetilde{\varphi}_{nlj}$ are the radial parts of the pseudospinors,
 $P_{nlj}$ and $Q_{nlj}$ are the radial parts of the large and small
 components of the Dirac spinors, and $\chi_{ljm}$ is the two-component
 spin-angular function.

 We shall assign an orbital to the OC or V subspaces using the average radius
 criterion rather than the orbital energy criterion. The difference in such
 a definition is essential mainly for transition metals, lanthanides and
 actinides in which orbitals with different principal quantum numbers may
 have close orbital energies. For the problems of the RECP generation and
 application, which we are going to discuss, the space criterion is more
 appropriate.

\section{GRECP operator}
 \label{S:GRECP}

 Following basically the scheme developed by K.~Pitzer's group (Lee {\it et
 al.}\ \cite{Lee}, Christiansen {\it et al.}\ \cite{Christ}), the numerical
 pseudospinors $\widetilde{\varphi}_{nlj}(r)$  are constructed of the large
 components $P_{nlj}(r)$ of the outer core and valence Dirac-Fock (DF) spinors
 so that the
 innermost pseudospinors of them (for each $l$ and $j$) are nodeless, the next
 pseudospinors have one node, and so forth:

\begin{equation}
 \widetilde{\varphi}_{nlj}(r) =
 \left\{
  \begin{array}{ll}
   P_{nlj}(r),                              &  r\geq R_{c}, \\
   f(r)=r^{\gamma}\sum_{i=0}^{5}a_{i}r^{i}, &  r<R_{c},
  \end{array}
 \right.
\end{equation}
\[
 \begin{array}{ccccc}
  l=0,1,\ldots,L, & & j=|l\pm\frac{1}{2}|, & & n=n_c,n_c',\ldots,n_v,
 \end{array}
\]
        where $L$ is one more than the highest orbital angular momentum
        of the IC spinors. The leading power $\gamma$ in
        the polynomial is typically chosen to be close to $L$ in order
        to ensure a sufficient ejection of the pseudovalence electrons
        from the IC region.

 To derive the GRECP components $U_{nlj}(r)$, the HF equations
 are inverted for the V and OC pseudospinors (for each
        $l=0,\ldots,L$ and $j=|l \pm \frac{1}{2}|$)
        so that $\widetilde{\varphi}_{nlj}$ are solutions of
	the nonrelativistic-type HF equations in the
	{\it jj}-coupling scheme for a ``pseudoatom'' with
        the removed IC electrons (Goddard~III, 1968)\ \cite{Goddard}
\begin{eqnarray}
 U_{nlj}(r) & = &\widetilde{\varphi}_{nlj}^{-1}(r)
		\biggl[\biggl(\frac{1}{2}\frac{d^{2}}{dr^{2}}
		+ \frac{1}{r}\frac{d}{dr}
		- \frac{l(l+1)}{2r^{2}}
		+ \frac{Z}{r} -
		\widetilde{\bf J}(r) +
		\widetilde{\bf K}(r)
		        \nonumber\\
	    & + & \varepsilon_{nlj}  \biggr) \widetilde{\varphi}_{nlj}(r) +
		\sum_{n'\neq n} \varepsilon_{n'nlj}
		\widetilde{\varphi}_{n'lj}(r) \biggr],
 \label{U_nlj}
\end{eqnarray}
        where $Z$ is the nuclear charge, $\widetilde{\bf J}$ and
        $\widetilde{\bf K}$ are the Coulomb and exchange operators calculated
        with the pseudospinors, $\varepsilon_{nlj}$ is the one-electron
        energy of the corresponding spinor, and $\varepsilon_{n'nlj}$ are
        the off-diagonal Lagrange multipliers.

 In general, different potentials $U_{nlj}$ are obtained for
 different pseudospinors $\widetilde{\varphi}_{nlj}$ with the help
 of Eq.~(\ref{U_nlj}).  It is necessary to
 construct such an RECP operator which would act on each individual
 pseudospinor as the corresponding potential.  This operator may be presented
 in the non-Hermitean (right-handed) form as:

\begin{equation}
  {\bf U}^{Ef}\ =\ \sum_n \sum_{l=0}^{\infty} \sum_{j=|l-1/2|}^{l+1/2}
            U_{nlj}(r) \widetilde{\bf P}_{nlj},
 \label{Oper_U1}
\end{equation}
  where

\[
\widetilde{\bf P}_{nlj} = \sum_{m=-j}^j
    \bigl| \widetilde{nljm} \bigl\rangle \bigr\langle \widetilde{nljm} \bigr|,
\]
\[
n=n_c,n_c',\ldots,n_v,n_a,n_a',\ldots,
\]
 i.e.,\ $|\widetilde{nljm}\rangle\langle \widetilde{nljm}|$ is the projector
 on the $\tilde{\psi}_{nljm}$  pseudospinor and the $n$ index runs over the
 OC, V, and A pseudospinor subspaces. The direct use of
 expression~(\ref{Oper_U1}) in calculations is impossible because it includes
 the infinite summations on the $n$ and $l$ indices.

 The $U_{nlj}$ potentials contain the contribution from the Coulomb and
 exchange interaction with the IC electrons, the contribution from smoothing
 the spinors, and the contribution from relativistic effects as a result of
 the replacement the Dirac Hamiltonian by the nonrelativistic-type
 Hamiltonian.

 With the help of the property

\begin{equation}
  \bigl\langle  \tilde{\varphi}_{nlj} \bigl| U_{nlj}-U_{n'lj} \bigr|
              \tilde{\varphi}_{n'lj} \bigr\rangle  = 0 ,
 \label{Property6}
\end{equation}
 when neglecting small terms, the operator (\ref{Oper_U1}) (Generalized RECP
 or GRECP operator) can be written in the Hermitian form:

\begin{eqnarray}
 {\bf U} & \simeq & U_{n_vLJ}(r)
                   + \sum_{l=0}^L \sum_{j=|l-1/2|}^{l+1/2}
                   \bigl[U_{n_vlj}(r)-U_{n_vLJ}(r)\bigr]
                   {\bf P}_{lj}                             \nonumber\\
             & + & \sum_{n_c} \sum_{l=0}^L \sum_{j=|l-1/2|}^{l+1/2}
                   \Bigl\{\bigl[U_{n_clj}(r)-U_{n_vlj}(r)\bigr]
                   \widetilde{\bf P}_{n_clj} + \widetilde{\bf P}_{n_clj}
                   \bigl[U_{n_clj}(r)-U_{n_vlj}(r)\bigr]\Bigr\}     \nonumber\\
             & - & \sum_{n_c,n_c'} \sum_{l=0}^L \sum_{j=|l-1/2|}^{l+1/2}
                   \widetilde{\bf P}_{n_clj}
                   \biggl[\frac{U_{n_clj}(r)+U_{n_c'lj}(r)}{2}-U_{n_vlj}(r)\biggr]
                   \widetilde{\bf P}_{n_c'lj},
 \label{UGRECP}
\end{eqnarray}
 where

\[
  {\bf P}_{lj} = \sum_{m=-j}^j
    \bigl| ljm \bigl\rangle \bigr\langle ljm \bigr|
\]
 is the ordinary spin-angular projector and the first line in
 Eq.~(\ref{UGRECP}) presents the standard radially local RECP operator.
 It can be shown\ \cite{GRThGr} that the action of this operator on the OC, V,
 and A pseudospinors is equivalent (with the required accuracy) to the action
 of the corresponding potentials on them.

 Note that the nonlocal terms with the projectors on the most important
 correlating functions $\tilde{\varphi}_{n_kl_kj}^{corr}(r)$ localized mainly
 in the OC and V regions and with the corresponding potentials,
 $U_{n_kl_kj}(r)$, in principle (i.e.\ in the cases in which it is essential),
 can be taken into account in the expression for the GRECP operator
 additionally to the terms with the OC projectors $\widetilde{\bf P}_{n_clj}$.

 With the help of the following identities for the ${\bf P}_{lj}$ projectors
 (Hafner \& Schwarz, 1979)\ \cite{Hafner}:

\begin{equation}
        {\bf  P}_{l,j=l\pm 1/2}(\Omega,\sigma)\
         =\ \frac{1}{2l+1} \Bigl[ \Bigl(l+\frac{1}{2}\pm \frac{1}{2}\Bigr)
           {\bf  P}_l(\Omega) \pm
          2 {\bf P}_l(\Omega)
          \vec{\bf l}\vec{\bf s} {\bf P}_l(\Omega) \Bigr] ,
\label{Oper_Pnljs}
\end{equation}
 the GRECP operator can be easily written in the spin-orbit representation.

\subsection{Calculations of Hg and Pb}
 \label{Calc-HgPb}

 The accuracy of the GRECP approximation was tested in correlation structure
 calculations of the Hg and Pb atoms in which we were able to use very
 flexible basis sets.
 The all-electron Fock-space Relativistic Coupled Cluster (RCC) method has been
 described in review~\cite{Kaldor} and its references.
 RECP errors in reproducing the all-electron RCC transition energies in Hg for
 the case of 20 correlated electrons and the equivalent correlation
 $[7,9,8,6,7,7]$ basis sets (for all-electron and RECP calculations,
 see~\cite{MosyaginHg2}) are presented in Table~\ref{RECP-Hg} (compiled
 from~\cite{MosyaginHg2}). The highest absolute error in reproducing the
 transition energies with excitation or ionization of single electron is
 94~cm$^{-1}$ for the GRECP~\cite{Tupitsyn1}, 729~cm$^{-1}$ for the RECP of
 Ross {\it et al.}~\cite{Ross} and 1747~cm$^{-1}$ for the energy-adjusted
 PP~\cite{Hausser}. The same number of electrons, twenty, is explicitly treated
 in all these RECP versions. The larger errors for RECPs~\cite{Ross} and
 \cite{Hausser} are mainly due to the neglect of the difference between the
 outer core and valence potentials in these RECP versions
 (see~\cite{Tupitsyn1,GRThGr} for more details).

 Calculations of transition energies in the Pb atom are performed for five
 lowest-lying states in~\cite{Isaev} (see Table~\ref{MBPT_22} here) by the
 four-component~\cite{Dzuba} and two-component (with GRECP)~\cite{MosyaginHg1}
 versions of the combined method of the Configuration Interaction (CI) and the
 second-order Many-Body Perturbation Theory (MBPT2)\ \cite{Dzuba}. The
 22e-GRECP version for Pb~\cite{Tupitsyn1} was tested and very good agreement
 of the GRECP/MBPT2/CI calculations with the four-component DC/MBPT2/CI results
 is obtained.

\section{Frozen core approximation for outer core shells}
 \label{OC-Fr}

 To perform precise calculations of the chemical and spectroscopic properties,
 the correlations should be taken into account not only within the valence
 regions of heavy atoms and heavy-atom molecules but within the core regions
 and between the valence and core electrons as well. In practice, the goal is
 to achieve a given level of accuracy correlating as small a number of
 electrons as possible, thus reducing the computational effort. However, the
 performance of the RECPs generated for a given number of explicitly treated
 electrons can not always satisfy the accuracy requirements of the
 correlation structure calculation.

 It can be illustrated on example of Tl~\cite{TitovTlH}.
 To attain a good accuracy, e.g., within 400 $cm^{-1}$ for a group of
 low-lying states of molecules containing Tl, one should correlate at least 13
 electrons of Tl (i.e.,\ including the $5d$ shell) and it would be optimal to
 use the RECPs with this number of electrons treated explicitly (i.e.,
 13e-RECP) like the RECP of Ross {\it et al.}~\cite{Ross} or our valence RECP
 version~\cite{Tupitsyn1}. However, all the known nodeless 13e-RECPs cannot
 ensure the above-mentioned accuracy and one should use the RECPs with at least
 21 electrons (e.g., 21e-GRECP for Tl of our group~\cite{Tupitsyn1}). The
 $5s$ and $5p$ pseudospinors can be treated as frozen, providing, nevertheless,
 the above-declared accuracy. The $5p$ orbitals have the orbital energies
 about four times higher and their average radii are 1.4 times shorter than
 those for the $5d$ orbitals. Moreover, their angular correlations are
 suppressed as compared with those for the $5d$ shell because the most
 ``profitable'' polarization states ($5d$ in this case) are completely occupied
 in the lowest-lying states.  Therefore, the $5s, 5p$ states are substantially
 less active in chemical processes.

 One can apply the energy level-shift technique in order to ``freeze'' the
 $5s$ and $5p$ pseudospinors~\cite{GRThGr,TitovTlH}. Following Huzinaga {\it et
 al.}~\cite{Huzinaga}, one should add the matrix elements of the SCF field
 operators (the Coulomb and spin-dependent exchange terms) with these OC
 pseudospinors to the one-electron part of the Hamiltonian together with the
 level-shift terms

\begin{equation}
  \sum_{n_{c_f},l,j}\  B_{n_{c_f}lj}\
        \sum_m \widetilde{| n_{c_f}l_{c_f}j m\rangle}
                           \widetilde{\langle n_{c_f}l_{c_f}j m |} ,
 \label{OC_Fr-1}
\end{equation}
 where $B_{n_{c_f}lj}$ is at least of order $|2\varepsilon_{n_{c_f}lj}|$ of
 magnitude and $\varepsilon_{n_{c_f}lj}$ are the orbital energies of the OC
 pseudospinors $\widetilde{\phi}_{n_{c_f}lj}(r)$ which are frozen. These
 nonlocal terms are needed in order to exclude a collapse of the molecular
 orbitals to the ``frozen'' states (the $5s_{1/2}, 5p_{1/2,3/2}$ pseudospinors
 for Tl).

 All terms with the frozen core pseudospinors described here (the Coulomb and
 exchange interactions, the level-shift operator) can easily be presented in
 the spin-orbit form with the help of Eq.~(\ref{Oper_Pnljs}), as was said above
 with respect to the GRECP operator (the spin-orbit form of the GRECP operator
 can be found, e.g., in~\cite{GRThGr}).
 More importantly, these OC pseudo{\it spinors} can be frozen (as spinors)
 in calculations with the {\it spin-orbit} basis sets and they can be
 frozen at the stage of calculation of the one-electron matrix elements of the
 Hamiltonian (as is implemented in the MOLGEP code~\cite{MOLGEP}). Thus,
 any integrals with the indices of the frozen spinors are completely
 excluded already after the integral calculation step.

\subsection{Calculation of TlH}
 \label{Calc-TlH}

 The Multi-Reference single- and Double-excitation Configuration Interaction
 (MRD-CI) method~\cite{BuenkerCI} with the Spin-Orbit (SO) configuration
 selection scheme~\cite{TitovTlH} was applied in the first precise GRECP
 calculation of molecules.
 In the GRECP/MRD-CI calculations of spectroscopic properties for the $0^+$
 ground state of TlH~\cite{TitovTlH} (see Table~\ref{TlH} here), the
 21e/8fs-GRECP for Tl was used (i.e., the 21 electron GRECP~\cite{Tupitsyn1}
 with 8 electrons occupying the frozen OC pseudospinors $5s_{1/2}$ and
 $5p_{1/2,3/2}$). The contracted [4,4,4,3,2] basis set for thallium generated
 in~\cite{TitovTlH} and the [4,3,1] basis for hydrogen (see~{\it
 http://www.\-qchem.pnpi.\-spb.ru}) contracted from the primitive (6,3,1)
 gaussian basis set of Dunning~\cite{Dunning} were used.

 Fourteen electrons are correlated in the calculation of spectroscopic
 constants in TlH and, as one can see from Table~\ref{TlH}, very good agreement
 is found with the experimental data contrary to other known calculations.
 Close results are obtained very recently in the GRECP/RCC calculations
 of TlH in our joint calculations with group of Prof.\ Kaldor (Tel-Aviv
 University). The GRECP/RCC and GRECP/MRD-CI calculations of HgH and HgH$^+$
 are in progress now.

\section{Correlations with inner core electrons and Breit effects}
 \label{C_Cor}

 On example of the mercury atom\ \cite{MosyaginHg2,MosyaginHg1} the importance
 of correlation of the inner core electrons (removed from RECP calculations)
 was investigated with the help of the combined MBPT2/CI method\ \cite{Dzuba}
 and with the RCC method\ \cite{Kaldor}. It is
 obtained that at least 34 external electrons (occupying the IC $4f$
 shell, outer core $5s, 5p$ and $5d$ shells, valence $6s, 6p$ etc.\ shells) of
 Hg should be correlated and the one-electron basis set should contain up to
 $h$ angular momentum functions in order to attain a reliable agreement with
 the experimental data for the low excitation energies within 100--200
 $cm^{-1}$, whereas the errors of the gaussian approximation of the GRECP
 components and the effects of different nuclear models are negligible for this
 accuracy.  Otherwise, e.g., for the 20 electron RECPs for Hg known from the
 literature, energies of excitations from the ground state can not be
 calculated with the accuracy better than 200--500 $cm^{-1}$.

 However, our test calculations show that the main contribution from the
 correlations with the $4f$ shell is due to the one-electron correction from
 the self-energy diagrams~\cite{Dzuba} and, therefore, this contribution can be
 taken into account for the 20 electron GRECP at the generation stage, i.e.\
 the correlations with the IC $4f$ shell can be described by the GRECP.
 The technique of generating the ``correlated'' GRECP is discussed by the
 authors in~\cite{GRThGr}. When using such GRECPs in calculations, both
 required basis sets for the corresponding atoms and computational efforts can
 be seriously reduced when performing correlation structure calculations.

 Turning back to the TlH calculation, further improvement of the accuracy can
 be attained when correlations with the outer core $4f,5p$ and $5s$ shells of
 Tl and Breit effects are taken into account. We expect that this can be
 efficiently done in the framework of the ``correlated'' 21e/8fs-GRECP version
 in which 13 electrons are treated explicitly as in the above considered
 GRECP/MRD-CI calculation of TlH.

 The first correlated GRECP we have generated for the Hg atom in the framework
 of the MBPT2 method. Despite the considerable improvement in accuracy with
 this GRECP version when calculating excitation energies in Hg, the
 reliability of the applied generation scheme was found to be not very high. A
 new scheme of generation of correlated GRECPs in the framework of the RCC
 method is in progress now.

 Breit and other QED effects are relatively large for lanthanides and
 actinides. The incorporation of these corrections to calculations with RECPs
 is discussed in our theoretical paper on the GRECP method~\cite{GRThGr}. Some
 recent test calculations have shown that the Breit and other QED corrections
 for outer core and valence electrons can be efficiently represented by a
 one-electron spin-dependent radially local operator and, thus, can be easily
 incorporated into GRECPs.

\section{RECPs for transition metals, lanthanides, and actinides}
 \label{SfC-section}

 The disadvantage of the commonly used RECP versions for the transition,
 rare-earth, etc., elements is that they still require the explicit treatment
 of essentially larger number of electrons as compared with other
 (``normal'') elements in order to attain the same computational accuracy.

 In paper~\cite{SfC-RECP}, we added ``self-consistent'' (SfC) terms to the
 RECP operator which take account of the circumstance that the RECPs
 generated for the different occupation numbers of the outermost
 $d$ or $f$ shells of these elements are somewhat distinguished.
 Below we will refer to these shells as OuterMost Core (OMC) shells following
 the partitioning made in subsection\ \ref{Notation}.
 The significant improvement was attained in reproducing the atomic
 excitation energies in comparison with the conventional RECP calculations.

 Whereas the form of the standard radially local RECP operator accounts for the
 property that the dependence of the corresponding potentials $U_{n_vlj}(r)$
 on the relaxation of the V spinors ($\varphi_{n_vlj}$) is weak in the bond
 making, the SfC correction to this operator takes into account the fact of a
 relatively small relaxation of the OC spinors $\varphi_{n_clj}$ caused by the
 change of the OMC shell occupation number in a majority of cases.


\vspace{.5cm}

 The main features of the SfC~RECP constructing for the case of the
 radially local RECP are:

\begin{enumerate}
\item   The all-electron DF calculations of two generator states with
        different occupation numbers of the OMC $d$ ($f$) shell are
        carried out for an atom under consideration.
        These occupation numbers are designated $N_1$ and $N_2$.

\item   The ``generating'' RECPs ${\bf U}^{N_1}$ and ${\bf U}^{N_2}$ are
        constructed for
        these generator states employing the standard scheme. The operator
        for a radially local RECP (incorporating the SfC corrections into
        the GRECP is made similarly) has the form

\begin{equation}
 {\bf U}^{N} = E_{core}^{N} + U_{LJ}^{N}(r) + \sum_{l=0}^{L}
                \sum_{j=|l-\frac{1}{2}|}^{l+\frac{1}{2}}
                            [U_{lj}^{N}(r)-U_{LJ}^{N}(r)]
                                          {\bf P}_{lj},
\end{equation}
        where $E_{core}^{N}$ is the core energy and $U_{lj}^{N}(r)$
        are the radial components of the RECP for each pair of quantum
        numbers $l$ and $j$. They are derived for the OMC $d$ ($f$) shell
        occupation number  $N$.

        This operator differs from the standard (``nodeless'') RECP operator
        by the $E_{core}^{N}$ constant. This value corresponds to the
        contribution of the core electrons (excluded from calculations with
        the conventional RECP) to the total atomic energy and it is
        usually excluded from the RECP calculations because the
        $E_{core}^{N}$ only shifts all the energy levels by a constant and
        does not influence on the chemical properties. In the case of the
        SfC~RECP, inequality $E_{core}^{N_1}\neq E_{core}^{N_2}$ should be
        taken into account when transitions with the change of the occupation
        number $N$ are under consideration.

\item   The one-electron SfC~RECP operator is written as

\begin{equation}
 {\bf U}^{SfC} = \frac{ {\bf U}^{N_{1}}+{\bf U}^{N_2} }{ 2 }
             + \frac{ {\bf U}^{N_{1}}-{\bf U}^{N_2} }{ N_1-N_2 }
             \biggl(N-\frac{N_{1}+N_{2}}{2}\biggr)
             + B \biggl(N - \frac{N_1+N_2}{2}\biggr)^2 ,
\label{SfC-oper}
\end{equation}
        where $B$ is the adjusting parameter and $N$ is the average value
        $\langle\Psi\mid {\bf N}\mid\Psi\rangle$
        of the OMC $d$ ($f$) shell occupation number operator which in
        the second quantization representation is

\begin{equation}
 {\bf N} = \sum_{jm}^{} \widetilde{\bf a}_{n_{omc}l_{omc}jm}^{\dagger}\
                     \widetilde{\bf a}_{n_{omc}l_{omc}jm} ,
\end{equation}
        where $n_{omc}$ and $l_{omc}$ are the principal and orbital quantum
        numbers of the OMC shell.

\end{enumerate}

 At every iteration of the MC~SCF, etc., procedures, the $N$ value can be
 determined for the wave function $\Psi$ without a serious complication of the
 calculation.

 We should add here that in a high-precision electronic structure calculation
 of a heavy-atom molecule, correlation with the electrons occupying the IC
 or/and frozen OC states (see section\ \ref{OC-Fr})
 can be substantial for calculated properties and dependent from the occupation
 numbers of states of V and other OC electrons as was demonstrated in\
 \cite{MosyaginHg2} for the OC-OC correlation of $5s$ and $5d$ electrons in Hg.
 Therefore, in principle, the SfC terms dependent on the occupation numbers of
 the V states can appear in the expression for the SfC RECP operator\
 (\ref{SfC-oper}). In this case, the use of the ``correlated'' RECPs ${\bf
 U}^{N_1}$ and ${\bf U}^{N_2}$ also can be required in\ (\ref{SfC-oper}).

\subsection{SfC~GRECP calculations of uranium}
 \label{SfC-GRECP-U}


 In Table~\ref{U_SfC1} (compiled from \cite{GRThGr}), one-configuration
 numerical calculations of transition and ionization energies in uranium are
 presented for our GRECP versions, for the RECP of Ermler {\it et al.}\
 \cite{Ermler1} and for the energy-adjusted pseudopotential of K\"uchle
 {\it et al.}\ \cite{Kuchle1} as compared with the DF energies.

 The GRECP correcting terms described in Section~\ref{S:GRECP} allows one to
 describe accurately the V region of uranium when the $5f$ shell occupation
 number close to that for the generator state. The addition of the SfC
 correction increases the accuracy of reproducing the energies of transitions
 without changing the $5f$ shell occupation number in two to eight
 times and in those with changing these occupation numbers in 1.5--3
 times. One can see from Table~\ref{U_SfC1} that the addition of the quadratic
 SfC (QSfC) correction (the last term in~(\ref{SfC-oper})) leads to a
 significant increasing (up to 20 times) of the SfC~GRECP accuracy for the
 transitions with the change of the OMC shell occupation number.



\section{Term-splitting correction}
 \label{TS_c}

 As one can see from Table~\ref{U_term1}, the above-discussed SfC corrections
 give no improvement in the term splitting of the $5f_{5/2}^3 6d_{3/2}^1
 7s_{1/2}^2$ configuration of uranium as compared with the generating
 GRECPs. Analysis of these errors shows that the main
 contribution (about 90 \%) is due to reproducing the OMC spinors with
 the help of the nodeless pseudospinors.

 Consider the GRECPs with 24 and 32 explicitly treated electrons for which the
 OMC $5f$ pseudospinors are nodeless and the $6d$ ones have radial nodes (for
 the RECPs with the 14-electron pseudovalence space, for which these OMC
 shells are described by the nodeless pseudospinors, the errors are greater in
 general and the analysis is more complicated).

 The simplest way to reduce drastically these errors is to generate and use
 such GRECP versions for which the $5f$ shell is characterized by nodal
 pseudospinors, whereas the $4f$ pseudospinors are nodeless. To reduce
 computational efforts, the $4f$ shell can be treated as frozen in the way
 described in section~\ref{OC-Fr}. However, when approximating the nodal $5f$
 pseudospinors, some additional basis functions and computational efforts will
 be required.

 If the small magnitude of the OMC shell ($5f$ here) relaxation is taken into
 account, there is another way out which can be applied for the cases of
 low-lying states. We suggest to add the terms

\begin{eqnarray}
  {\bf U}^{TSc} &=& \sum_{x_1,x_2,x_3,x_4}  \lambda_{x_1x_2,x_3x_4}
   \widetilde{\bigl| x_1 \bigl\rangle} \widetilde{\bigr\langle x_2 \bigr|}
  \cdot
   \widetilde{\bigl| x_3 \bigl\rangle} \widetilde{\bigr\langle x_4 \bigr|}
 \nonumber\\
 &-&2  \sum_w \sum_{x_1,x_2,x_3}  (\lambda_{x_1x_2,x_3x_3}-
 \lambda_{x_1x_3,x_3x_2}) \delta_{wx_3}
   \widetilde{\bigl| x_1 \bigl\rangle} \widetilde{\bigr\langle x_2 \bigr|} ,
 \label{TS-c1}
\end{eqnarray}
 where
$
   x_i = (n_{omc},l_{omc},j_i,m_i)
$
 \footnote{
   The $j_i$ and $m_i$ indices run over all the possible values for a given
   ($n_{omc},l_{omc}$).
   In the case of the partially occupied OMC $5f$ shell, the $x_i$ indices
   correspond to both OC and virtual pseudospinors (i.e., occupied and
   unoccupied one-electron states in the generator state).
 },
 to the GRECP operator. These terms are correcting the one- and two-electron
 integrals with the $5f$ pseudospinors (and only these integrals).

 The results of our calculations (Table~\ref{U_term1}) show that the addition
 of the Term-Splitting (TS) correction (called in~\cite{GRThGr} by the SO
 correction) allows one to reduce the most serious errors about 10 times for
 the term splittings, thus reducing the errors for the transition energies
 between terms to the same order of magnitude as are the errors for transitions
 between the states averaged over the different configurations
 (Table~\ref{U_SfC1}).

\section{Electronic densities and singular operators in atomic cores}
 \label{Rest}

 To evaluate the matrix elements of the operators being singular in the core
 regions of heavy atoms in a molecule after the RECP calculation of this
 molecule at the first step, the proper shapes of the valence four-component
 molecular spinors must be restored in the atomic core regions at the second
 step.

 The applicability of the two-step algorithms to the electronic structure
 calculation of heavy-atom molecules is a consequence of the fact that the
 valence and core electron subspaces may be represented as the subsystems,
 interaction between which is described mainly by some integrated and not by
 detailed properties of these subsystems. As a result, the use of the GRECP
 allows one to reproduce accurately the original electronic densities in the
 valence region for excited states as it is demonstrated on U in~\cite{GRThGr}
 and on Hg, Tl, Pb, and Bi  in~\cite{Tupitsyn1}.  Another result is that the
 electronic structure can be restored in the core regions of heavy atoms after
 the GRECP calculation of a molecule with a good accuracy.

 Both the nonvariational procedures (Pacios \& Christiansen, 1985; Titov,
 1986)\ \cite{Pacios1,Titov0} and the variational technique (Titov, 1996)\
 \cite{TitovVRest} for the electronic structure restoration after the
 shape-consistent RECP calculation were suggested.

 In~\cite{Titov0,PbFHgF,TitovYbF,KozlovBaF,MosyaginYbF}, the nonvariational
 restoration technique was applied to evaluation of the hyperfine constants and
 the parity non-conservation effects in the PbF, HgF, BaF, and YbF molecules.
 The molecular pseudoorbitals in the form of linear combinations of atomic
 pseudoorbitals were evaluated at the RECP calculation stage. Then, the atomic
 pseudoorbitals were replaced by the unsmoothed four-component DF spinors
 derived for the same atomic configurations as the basis pseudoorbitals. The
 MO~LCAO coefficients were preserved after the RECP calculation. Such
 constructed molecular relativistic spinors were used for calculation of the
 P,T-odd spin-rotational Hamiltonian parameters.

 In a general case, the first-order reduced density matrix,
 $\{\widetilde{D_{pq}}\}$, should be evaluated after the molecular RECP
 calculation on the basis set of atomic pseudospinors whereas the matrix
 elements $\{A_{pq}^K\}$ of the one-electron
 operators of hyperfine interaction etc., $\{{\bf A}^K\}$, should be evaluated
 in the equivalent basis set of atomic four-component spinors. Then the
 physical values are calculated as

\begin{equation}
    \langle {\bf A}^K \rangle\ =\ \sum_{pq} \widetilde{D_{pq}} A_{pq}^K\ .
 \label{<A^K>}
\end{equation}

 The latest {\it ab initio} calculations of the hyperfine, $P$-odd, and
 $P,T$-odd constants for the BaF and YbF molecules are performed by our group
 with the help of the GRECP/RASSCF/EO scheme~\cite{KozlovBaF,MosyaginYbF}
 (RASSCF is the Restricted Active Space SCF method~\cite{RASSCF} and EO is
 Effective Operator technique based on MBPT2~\cite{Dzuba,KozlovBaF}). This
 technique allows one to take into account polarization and correlation in the
 valence and core regions. The electronic wave function of BaF and YbF for the
 ground ($^2\Sigma_{1/2}$) states is calculated in the GRECP approximation. The
 molecular four-component spinors in the core region of heavy atoms are
 restored in the framework of the non-variational procedure. Core
 polarization/correlation effects are taken into account with the help of the
 atomic MBPT2 when constructing the effective valence Hamiltonians for the Ba
 and Yb atoms.

\subsection{Calculations of YbF}
 \label{YbF-rest}

 The results for the YbF molecule by Mosyagin {\it et al.}
 (1998)~\cite{MosyaginYbF} are presented in Table~\ref{P,T-odd} compared with
 other results. For the isotropic hyperfine constant
 $A = (A_{\parallel}+2A_{\perp})/3$, the accuracy of our calculation is about
 3\% as compared to the experimental datum. The dipole constant $A_{\rm d} =
 (A_{\parallel}-A_{\perp})/3$ (which is much smaller in magnitude), though
 better than in all previous calculations known from the literature, is still
 underestimated by almost 23\%. Being corrected within a semiempirical approach
 for a perturbation of $4f$-shell in the core of Yb due to the bond making,
 this error is reduced to 8\%. Our value for the effective electric field on
 the unpaired electron is 4.9 a.u.=$2.5 \times 10^{10}$ V~cm$^{-1}$.

 One can see from Table~\ref{P,T-odd} that the values of the $W_d$ constant
\[
        W_d d_{\rm e} =
        2 \langle ^2\Sigma_{1/2}|H_{d}| ^2\Sigma_{1/2} \rangle ,
\]
 where $H_{d}$ describes interaction of the electron EDM (i.e.\ Electric
 Dipole Moment) $d_{\rm e}$ with the internal molecular field
 ${\bf E}^{\rm mol}$:

\begin{eqnarray}
        &&H_{d} = 2 d_{\rm e}
        \left(\begin{array}{cc}
        0 & 0 \\
        0 & {\bf \sigma}
        \end{array} \right)
        \cdot {\bf E}^{\rm mol},
\end{eqnarray}
 from the unrestricted DF calculation of Parpia (1998)~\cite{Parpia}, the most
 recent semiempirical calculation of Kozlov (1997)~\cite{Kozlov1YbF} and our
 latest GRECP/RASSCF/EO calculation are in a very close agreement now. It
 should be noted that the valence electron contribution to the $W_d$
 in~\cite{Parpia} is in  7.4\%  agreement with the corresponding RECP-based
 calculation of Titov {\it et al.} (1996)~\cite{TitovYbF}. Another recent DF
 calculation of Quiney {\it et al.} (1998)~\cite{Quiney} gives the value, which
 is two times smaller.

 It is clear from the calculation that the $4f$ electrons in the core of
 ytterbium should be explicitly treated in the planned GRECP calculation of YbF
 by the relativistic coupled cluster method
 in order to obtain the parameters of $P,T$-odd spin-rotational Hamiltonian
 with better accuracy.

\subsection{Variational restoration}
 \label{Var-rest}

 In the variational technique of the restoration\ \cite{TitovVRest}, the
 proper behavior of the four-component molecular spinors in the core regions of
 heavy atoms may be restored as an expansion on the spherical harmonics inside
 the sphere with an atomic core radius $R_{rest}$. The outer parts of spinors
 are frozen after the RECP calculation of the molecule considered. This method
 enables one to combine the advantages of two well-developed approaches:
 molecular RECP calculations (with gaussian basis sets) and one-center
 calculations (of atomic type with, e.g., numerical functions) in the most
 optimal way.

 The most promising application of the two-step method is in a
 possibility to ``split'' the correlation structure calculation of a molecule
 on two consequent calculations in the valence and core regions when the OC
 and V electrons are explicitly involved into the GRECP calculation and the
 OC and IC space regions are treated at the one-center restoration stage.

 Roughly speaking, the computational efforts for the correlation structure
 calculations in the core and valence regions are added within the two-step
 technique, whereas in the ordinary one-step scheme, they are multiplied by
 each other.

 The efficiency of the two-step technique was confirmed in the BaF and YbF
 calculations\ \cite{KozlovBaF,MosyaginYbF}. Even these not yet perfect
 calculations have provided a high accuracy (5--10 \%) for the P,T-odd
 spin-rotational Hamiltonian parameters (which can be compared with the 50 \%
 error for the hyperfine constants in YbF with the correlations in the V region
 taken into account within RASSCF), thus confirming a considerable promise of
 the two-step scheme for the correlation structure calculations.

 \acknowledgments

 This work was supported by the INTAS grant N 96-1266, the DFG/RFBR grant N
 96--03--00069, and the RFBR grant N 99--03--33249.
%
 The most consuming molecular calculations were carried out at the computer
 centers of the Bergische Universit{\"a}t GH Wuppertal and Tel Aviv university
 with the help of the JECS codes \cite{JECS}.




\begin{table}
\caption{ Transition energies by the 20e-RCC calculations of Hg with different
          RECP versions in the $[7,9,8,6,7,7]$ correlation basis set as
          compared with all-electron calculations. All values are in $cm^{-1}$.}

\vspace{1cm}

\begin{tabular}{@{}lrrrrrrrr}
 State                     & &\multicolumn{2}{c}{All-el.\ }
                                             & &\multicolumn{2}{c}{20 el.\ }
                                                               &20 el.\ &energy- \\
\cline{3-4}
 (Leading                  & & finite & point  & &\multicolumn{2}{c}{GRECP$^{\rm a}$}
                                                               &RECP of &adjusted\\
\cline{6-7}
 configuration,            & & nucl.\ & nucl.\ & &  num.\ & gaus.\ &Ross et al.$^{\rm b}$
                                                                        &20 el.\ PP$^{\rm c}$\\
\cline{3-9}
 Term)                     & &\multicolumn{2}{c}{Transition energies}& &\multicolumn{4}{c}{Errors$^{\rm d}$}\\
\cline{1-1}
\cline{3-4}
\cline{6-9}
 \multicolumn{9}{l}{$5d^{10} 6s^2 (^1S_0) \rightarrow$}                            \\
 $5d^{10} 6s^1 6p^1 (^3P_0)$ & &  37197 &  37260 & &    -16 &   -10  &   363  &  -666  \\
 $5d^{10} 6s^1 6p^1 (^3P_1)$ & &  39029 &  39091 & &     -1 &     4  &   378  &  -416  \\
 $5d^{10} 6s^1 6p^1 (^3P_2)$ & &  43860 &  43925 & &     22 &    28  &   437  &   619  \\
 $5d^{10} 6s^1 6p^1 (^1P_1)$ & &  55036 &  55095 & &     78 &    79  &   416  &   229  \\
 $5d^{10} 6s^1 (^2S_{1/2})$  & &  84159 &  84215 & &     28 &    36  &   482  &   -19  \\
 \multicolumn{9}{l}{$5d^{10} 6s^1 (^2S_{1/2}) \rightarrow$}                        \\
 $5d^{10} 6p^1 (^2P_{1/2})$  & &  51656 &  51734 & &    -11 &    -9  &   428  &  -796  \\
 $5d^{10} 6p^1 (^2P_{3/2})$  & &  60802 &  60883 & &      9 &    11  &   539  &   951  \\
 $5d^{10} (^1S_0)$           & & 150636 & 150720 & &    -67 &   -58  &   729  &   112  \\
\end{tabular}

\vspace{.5cm}

\noindent $^{\rm a}$The GRECP from reference~\cite{Tupitsyn1} with the
                    potentials taken in the numerical and gaussian forms,
                    respectively.

\noindent $^{\rm b}$The RECP from reference~\cite{Ross}.

\noindent $^{\rm c}$The PP from reference~\cite{Hausser} with the corrected
    $V_{so}$ by the factors $(2l+1)/2$ (M.\ Dolg, private communication).

\noindent $^{\rm d}$In this table, errors were
                    calculated as differences between the transition energies
                    from the RECP and all-electron calculations for the same
                    number of correlated electrons and equivalent
                    basis sets. The point nuclear model was employed.
\label{RECP-Hg}
\end{table}


\begin{table}
\begin{center}
\caption{ Absolute errors (AE) of Dirac-Coulomb (DC) 4e-CI and
          14e,22e-MBPT2/CI calculations in reproducing the experimental
          transition energies (TE) for the lowest-lying states of Pb (AE exp.);
          AE of the 22e-GRECP in reproducing DC calculations (AE DC) in $cm^{-1}$.
          Basis sets are $[3,5,3,2]$ for CI and $[4,5,4,2,2]$ for MBPT2. }

\vspace{1cm}

\begin{tabular}{lccrrrrrrr}
         & & & &           \multicolumn{2}{c}{4e-Cl} &
                                    \multicolumn{2}{c}{14e-MBPT2/CI} &
                                                     \multicolumn{2}{c}{22e-MBPT2/CI}\\
\cline{5-10}
 Conf.   &  Term   &  J  &Experiment &~~~DC~~& GRECP &~~~DC~~& GRECP &~~~DC~~& GRECP \\
\hline                                                               
         &         &     &        TE &AE exp.& AE DC &AE exp.& AE DC &AE exp.& AE DC \\
\cline{4-10}                                                          
 $6p^2$  &  $^3$P  &  0  &         0 &     0 &     0 &     0 &     0 &     0 &     0 \\
 $6p^2$  &  $^3$P  &  1  &      7819 &  -807 &  (26) &  -535 &  (56) &  -393 &  (69) \\
 $6p^2$  &  $^3$P  &  2  &     10650 &  -752 &  (36) &  -428 &  (62) &  -294 &  (77) \\
 $6p^2$  &  $^1$D  &  2  &     21458 & -1707 &  (46) &  -849 &  (89) &  -573 & (120) \\
 $6p^2$  &  $^1$S  &  0  &     29467 & -1553 &  (57) &  -270 &  (90) &   -33 & (127) \\
\end{tabular}
\label{MBPT_22}
\end{center}
\end{table}


\begin{table}
\begin{center}
\caption{ GRECP/MRD-CI and other calculations of the spectroscopic constants
          for the ground state of TlH. }

\vspace{1cm}

\begin{tabular}{lclc}
                                               & $R_e$ &$\omega_e$ &$D_e$  \\
Method                                         &($\AA$)&($cm^{-1}$)&($eV$) \\
\hline
SOCIEX: Tl [8,8,5,2] + H [4,3,1]               &       &           &       \\
(Rakowitz \& Marian, 1997 \cite{RakowitzTlH})  & 1.86~ &    1386   & 2.13~ \\
\hline
13e-RECP/SOCI: Tl [4,4,4,1] + H [4,2]          &       &           &       \\
(DiLabio \& Christiansen, 1998 \cite{DiLabio}) & 1.912 &    1341   & 1.908 \\
\hline
REP-KRCCSD[T]: Tl [4,5,5,1] + H [3,2]          &       &           &       \\
(Lee {\it et al.}, 1998 \cite{HSLee})          & 1.910 &    1360   & 2.02~ \\
21e-REP/KRCCSD(T): Tl [4,5,5,1] + H [3,2]      &       &           &       \\
(Han {\it et al.}, 2000 \cite{YKHan})          & 1.877 &           & 2.00~ \\
\hline
21e/8fs-GRECP/14e-MRD-CI Tl [4,4,4,3,2] + H [4,3,1] &  &           &       \\
(Titov {\it et al.}, 2000 \cite{TitovTlH})     & 1.870 &    1420   & 2.049 \\
\hline
Experiment (Grundstr\"om \& Valberg, 1938
                             \cite{Grundstr})  & 1.866$^{\rm a}$
                                                       &    1390.7 & 2.06~ \\
Experiment (Urban {\it et al.}, 1989 
                                \cite{Urban})  & 1.872$^{\rm b}$
                                                       &    1391.3 &       \\
\end{tabular}
\label{TlH}
\end{center}

\noindent $^{\rm a}$Huber \& Herzberg (1979) \cite{Huber} have published value
                   1.87 $\AA$ which can be obtained from the rotational
                   constant $B_e$.

\noindent $^{\rm b}$This value is calculated by us from $B_e$.

\end{table}


\begin{table}
\begin{center}
\caption{ Reproducing the all-electron transition energies (in $cm^{-1}$)
          between states of uranium (averaged over nonrelativistic
          configurations) in one-configuration calculations with RECPs. }


\begin{tabular}{lrrrrrrrrr}
                              &          &   RECP of  &  Energy-   &            & Quadratic  &\multicolumn{2}{c}{``Frozen}& \\
                              &   DF     &   Ermler   &  adjusted  & SfC        & SfC        &\multicolumn{2}{c}{  core''}& \\
                              &          &{\it et al.}~\cite{Ermler1}
                                         &   PP~\cite{Kuchle1}     & GRECP      & GRECP      &    ($f^3$) &   ($f^2$)  &    \\
\cline{2-8}
Num.\  of el-ns               &  All     &    14      &    32      &    24      &    24      &    24      &    24      &    \\
\cline{2-8}
       Conf.\                 &Tr.energy &                           \multicolumn{6}{c}{Absolute  error}                              \\
\hline
$5f^3 7s^2 6d^1 \rightarrow $ &          &            &            &            &            &            &            &    \\
$5f^3 7s^2 7p^1             $ &    7383  &        387 &       -498 &        -35 &        -33 &          2 &         14 &    \\
$5f^3 7s^2                  $ &   36159  &        332 &        130 &          4 &          6 &          3 &         16 &    \\
$5f^3 7s^1 6d^2             $ &   13299  &       -192 &       -154 &         -3 &         -5 &         -1 &        -16 &    \\
$5f^3 7s^1 6d^1 7p^1        $ &   17289  &        144 &       -621 &        -31 &        -31 &         -1 &         -5 &    \\
$5f^3 7s^1 6d^1             $ &   42436  &         98 &       -188 &        -18 &        -17 &          0 &         -5 &    \\
$5f^3 6d^2                  $ &   54892  &       -121 &       -398 &        -14 &        -15 &          1 &        -21 &    \\
$5f^3 7s^2 6d^1 \rightarrow $ &          &            &            &            &            &            &            &    \\
$5f^4 7s^2                  $ &   16483  &        176 &        788 &       -723 &          0 &         54 &        187 &    \\
$5f^4 7s^2      \rightarrow $ &          &            &            &            &            &            &            &    \\
$5f^4 7s^1 6d^1             $ &   15132  &       -738 &        -87 &         11 &        -11 &        -16 &        -35 &    \\
$5f^4 7s^1 7p^1             $ &   15016  &         90 &       -443 &        -37 &        -26 &         -1 &         -2 &    \\
$5f^4 7s^1                  $ &   38913  &         82 &       -110 &        -37 &        -22 &          1 &          3 &    \\
$5f^4 6d^2                  $ &   34022  &      -1287 &       -153 &         28 &        -13 &        -26 &        -62 &    \\
$5f^4 6d^1 7p^1             $ &   32341  &       -794 &       -457 &        -11 &        -23 &        -17 &        -39 &    \\
$5f^4 6d^1                  $ &   53637  &       -874 &       -245 &        -21 &        -29 &        -17 &        -39 &    \\
$5f^3 7s^2 6d^1 \rightarrow $ &          &            &            &            &            &            &            &    \\
$5f^2 7s^2 6d^2             $ &    3774  &       3096 &       -748 &        -17 &        -17 &         90 &        -96 &    \\
$5f^2 7s^2 6d^2 \rightarrow $ &          &            &            &            &            &            &            &    \\
$5f^2 7s^2 6d^1 7p^1        $ &   12646  &       -441 &       -626 &        -16 &        -15 &         -5 &          0 &    \\
$5f^2 7s^2 6d^1             $ &   42638  &       -498 &        155 &         24 &         25 &         -5 &          1 &    \\
$5f^2 7s^1 6d^3             $ &   10697  &        608 &       -240 &        -10 &        -10 &         13 &          1 &    \\
$5f^2 7s^1 6d^2 7p^1        $ &   19319  &        390 &       -826 &        -26 &        -26 &          6 &          0 &    \\
$5f^2 7s^1 6d^2             $ &   45478  &        402 &       -279 &        -13 &        -13 &          6 &          0 &    \\
$5f^2 6d^3                  $ &   54986  &       1127 &       -581 &        -14 &        -15 &         22 &          3 &    \\
$5f^3 7s^2 6d^1 \rightarrow $ &          &            &            &            &            &            &            &    \\
$5f^1 7s^2 6d^3             $ &   29597  &      11666 &      -1526 &       -896 &       -104 &        466 &         48 &    \\
$5f^1 7s^2 6d^3 \rightarrow $ &          &            &            &            &            &            &            &    \\
$5f^1 7s^2 6d^2 7p^1        $ &   18141  &      -1367 &       -778 &         46 &         49 &         -2 &         -2 &    \\
$5f^1 7s^2 6d^2             $ &   49158  &      -1355 &        173 &         70 &         73 &         -3 &         -2 &    \\
$5f^1 7s^1 6d^4             $ &    7584  &       1655 &       -331 &        -39 &        -40 &         22 &         14 &    \\
$5f^1 7s^1 6d^3 7p^1        $ &   21154  &        779 &      -1055 &        -11 &        -11 &         16 &         10 &    \\
$5f^1 7s^1 6d^3             $ &   48146  &        909 &       -381 &        -13 &        -13 &         17 &         10 &    \\
$5f^1 6d^4                  $ &   54235  &       2810 &       -782 &        -43 &        -45 &         42 &         27 &    \\
$5f^3 7s^2 6d^1 \rightarrow $ &          &            &            &            &            &            &            &    \\
$5f^5                       $ &  100840  &        430 &       1453 &      -1860 &         22 &        105 &        291 &    \\
\end{tabular}
\label{U_SfC1}
\end{center}
\end{table}


\begin{table}
\begin{center}
\caption{ Reproducing the all-electron transition energies between terms for
          different configurations of uranium (in $cm^{-1}$) in
          one-configuration calculations with GRECPs. }

\vspace{1cm}

\begin{tabular}{lrrrrr}
                     &          &            &            &            &            \\
                     &          &            &  TS-corr.\ &            &  TS-corr.\ \\
                     &          &     SfC    &     SfC    &            &            \\
                     &   DF     &    GRECP   &    GRECP   &   GRECP    &   GRECP    \\
\cline{2-6}
Num.\  of el-ns      &   all    &    24      &    24      &    32      &    32      \\
\cline{2-6}
   Conf.,            &Trans.\   & \multicolumn{4}{c}{Absolute}                      \\
       Term          &  energy  & \multicolumn{4}{c}{error}                         \\
\hline
                     &          &            &            &            &            \\
\multicolumn{6}{l}{ $5f_{5/2}^3 6d_{3/2}^1 7s_{1/2}^2 \;\;\;\;\; J=0 \rightarrow $ }\\
 $ J=1             $ &   18643  &         86 &        -97 &        142 &        -14 \\
 $ J=2             $ &    9729  &        136 &        -56 &        146 &        -13 \\
 $ J=3             $ &    7813  &       -117 &        -44 &        -69 &         -8 \\
 $ J=4             $ &    6759  &       -140 &        -36 &        -94 &         -5 \\
 $ J=5             $ &   -7918  &       -547 &         42 &       -485 &        -11 \\
 $ J=6             $ &  -10695  &       -509 &         76 &       -464 &        -31 \\
                     &          &            &            &            &            \\
\multicolumn{6}{l}{ $5f_{5/2}^3 5f_{7/2}^1 7s_{1/2}^2 \;\;\;\;\; J=1 \rightarrow $ }\\
 $ J=2             $ &    4392  &        175 &        -52 &        150 &        -39 \\
 $ J=3             $ &    2843  &        113 &        -33 &         97 &        -24 \\
 $ J=4             $ &    3477  &        133 &        -40 &        113 &        -30 \\
 $ J=5             $ &    2805  &        105 &        -32 &         88 &        -23 \\
 $ J=6             $ &    4631  &        169 &        -56 &        138 &        -42 \\
 $ J=7             $ &   -5951  &       -275 &         48 &       -253 &         28 \\
 $ J=8             $ &   -5450  &       -258 &         42 &       -239 &         24 \\
                     &          &            &            &            &            \\
\multicolumn{6}{l}{ $5f_{5/2}^2 6d_{3/2}^2 7s_{1/2}^2 \;\;\;\;\; J=0 \rightarrow $ }\\
 $ J=1             $ &  -19078  &       -487 &         -7 &       -461 &        -64 \\
 $ J=2             $ &  -15304  &       -331 &          6 &       -326 &        -45 \\
 $ J=3             $ &  -23607  &       -682 &         -8 &       -644 &        -85 \\
 $ J=4             $ &  -25984  &       -715 &         24 &       -689 &        -74 \\
 $ J=5             $ &  -32497  &       -848 &         19 &       -818 &        -94 \\
 $ J=6             $ &  -39551  &       -785 &         82 &       -788 &        -62 \\
                     &          &            &            &            &
\end{tabular}
\label{U_term1}
\end{center}
\end{table}


\begin{table}
\begin{center}
\caption{ Parameters of the spin-rotational Hamiltonian for $^{171}$YbF.
 }


\begin{tabular}{lcrlrr}
                             &   $A$   &$A_{\rm d}$& $W_d$    & $W_{\rm A}$
                             & $W_{\rm S}$ \\
Method                       & (MHz)   &  (MHz)
                             & ($10^{25}~\frac{\rm Hz}{\rm e\cdot cm}$)
                                                              & (Hz)  & (kHz) \\
\hline                       
Semiemp. (Kozlov {\it et al.},
      1994 \cite{KozlovYbF}) &         &           & $-$1.5   &  730  & $-$48 \\
Semiemp. (Kozlov, 1997
          \cite{Kozlov1YbF}) &         &           & $-$1.26  &       & $-$43 \\
\hline                       
GRECP/SCF (Titov {\it et al.},
       1996 \cite{TitovYbF}) &  4932   &   59      & $-$0.91  &  484  & $-$33 \\
GRECP/RASSCF                 &  4854   &   60      & $-$0.91  &  486  & $-$33 \\
\hline                       
DHF      (Quiney {\it et al.},
         1998 \cite{Quiney}) &  5918   &   35      & $-$0.31  &  163  & $-$11 \\
DHF+CP                       &  7865   &   60      & $-$0.60  &  310  & $-$21 \\
DHF (rescaled)               &         &           & $-$0.62  &  326  & $-$22 \\
\hline                       
Unrestricted DF (Parpia, 1998
              \cite{Parpia}) &         &           & $-$1.203 &       & $-$22 \\
DF (unpaired elect-          &         &           &          &       &       \\
~~~ron contribution)         &         &           & $-$0.962 &       &       \\
\hline                       
GRECP/RASSCF/EO              &         &           &          &       &       \\
 (Mosyagin {\it et al.}, 1998
         \cite{MosyaginYbF}) &  7842   &   79      & $-$1.206 &  634  &       \\
GRECP/RASSCF/EO              &         &           &          &       &       \\
~~~(with $4f$-hole           
   correction)               &  7839   &   94      & $-$1.206 &  634  &       \\
\hline                       
Experiment (Knight {\it et al.}, 
          1970 \cite{Knight} &  7617   &  102      &          &       &       \\
\end{tabular}
\label{P,T-odd}
\end{center}
\end{table}

\end{document}